# Performance Comparison of Cooperative and Distributed Spectrum Sensing in Cognitive Radio

Zheng SUN, Wenjun XU, Zhiqiang HE and Kai NIU
School of Information Engineering, Beijing University of Posts and Telecommunications, Beijing, China
zhengs.bupt@gmail.com

***Abstract*—In this paper, we compare the performances of cooperative and distributed spectrum sensing in wireless sensor networks. After introducing the basic problem, we describe two strategies: 1) a cooperative sensing strategy, which takes advantage of cooperation diversity gain to increase probability of detection and 2) a distributed sensing strategy, which by passing the results in an inter-node manner increases energy efficiency and fairness among nodes. Then, we compare the performances of the strategies in terms of three criteria: agility, energy efficiency, and robustness against SNR changes, and summarize the comparison. It shows that: 1) the non-cooperative strategy has the best fairness of energy consumption, 2) the cooperative strategy leads to the best agility, and 3) the distributed strategy leads to the lowest energy consumption and the best robustness against SNR changes.**

## I. Introduction

Recently the field of *cognitive radio* (CR) has drawn great interest, since this novel technology provides promising solution to enhance the spectrum efficiency of today's wireless network. Studies have shown that spectrum is extremely underutilized [1]. One way to increase the utilization is to design CR networks, where wireless equipments use smart radio to detect temporal and spatial "holes" in the spectrum, thus learn from the environment and perform further functions to serve the users.

A significant feature of CR networks is to allow users to operate in licensed bands without a license. However, since CR has to limit its interference to the primary network, CR users using a licensed band must vacate the band due to the presence of the primary user. Thus it is significant to detect the presence of licensed (primary) users by spectrum sensing in a very short time. Recent work considers how to take advantage of the local oscillator leakage power emitted from RF receivers to allow cognitive radios to sense and locate the primary users [2]. Some physical layer issues of spectrum sensing are discussed in [3]. For radio sensitivity of the sensing function through processing gain, the authors of [4] study three digital signal processing techniques.

In this paper, we are going to discuss how to deal with spectrum sensing in wireless sensor networks (WSN). Related work includes [2], which gives a physical layer and MAC layer solution of sensor nodes but lacks further designing on network architecture. Rather, we will discuss two strategies for efficient spectrum sensing in WSN. The first is cooperative sensing. Cooperative techniques are widely studied recently ([5]-[8]) to achieve a new form of spatial diversity via the cooperation of users [5]. In [6], the authors study two-user cooperative spectrum sensing in cognitive radio and show that, by allowing the cognitive users operating in the same band to cooperate, the detection time reduces and thus the overall agility increases. In [7], light-weight cooperation in sensing based on hard decision is proposed to mitigate the sensitivity requirements on individual radios. And in [8], the cooperative situations are considered by using game theory, and the authors show how the lack of cooperation affects the performance. In this paper, the cooperative sensing is described in a multi-node WSN network, in which multi-user diversity gain is further achieved.

The second strategy being discussed is distributed spectrum sensing, which is specialized from distributed learning and estimation theory [10]. To our best knowledge, distributed spectrum sensing is a rather fresh topic. The reason of adopting distributed spectrum sensing in WSN is twofold. Firstly, traditional cooperation in WSN needs *node* $\rightarrow$ *fusion center* transmissions with length of $O(1)$, while distributed sensing strategy adopts inter-node transmissions with length of only $O\left(\sqrt{\log^2 n/n}\right)$ [9], which therefore reduces energy costs and increases overall network longevity. Secondly, in this paper we show that by using distributed sensing strategy, the probability of detection at the fusion center is greatly increased comparing with both non-cooperative and cooperative sensing strategy.

The main purpose of this paper is to present strategies of cooperative and distributed spectrum sensing in WSN, and to compare their performances in terms of agility, energy efficiency, and robustness against SNR changes. The results drawn here may act as a reference for further researches.

The rest of the paper is organized as follows. In Section II, we describe the basic problem and a non-cooperative spectrum sensing strategy as a baseline. In Section III, a cooperative strategy utilizing inter-node information is discussed. And in Section IV we use distributed estimation theory to develop a distributed spectrum sensing strategy. Then in Section V, we discuss and compare the performances of the three strategies. And Section VI gives a summary and concludes the work.

This research is sponsored by Project 60772108 supported by National Natural Science Foundation of China and National Basic Research Program of China (973 Program), 2007CB310604

## II. NON-COOPERATIVE SPECTRUM SENSING

In this section, we describe the basic spectrum sensing problem in WSN and a spectrum sensing strategy without utilizing cooperation among sensor nodes. This strategy will be used as a baseline of the other two strategies throughout the paper.

### A. *Basic Detection Problem*

Let us consider a WSN with $N$ nodes operating in a TDMA mode. Assume that the nodes are uniformly distributed over a square with side of unit length as illustrated in Figure 1. Each node measures and therefore senses the presence of the primary user independently, and then transmits its detection results to the fusion center. The signal received by every sensor node is given by

$$y_i = P \cdot \theta \cdot h_{pi} + w_i \, , \tag{1}$$

where $P$ is the transmit power of the primary user, $h_{pi}$ denotes the instantaneous channel gain between the primary user and the $i$th node, and $w_i$ denotes the additive Gaussian noise. We assume that $h_{pi}$ and $w_i$ are independent zero-mean complex Gaussian random variables with variances $\sigma^2{}_h$ and $\sigma^2{}_w$ , respectively. Also, $\theta$ is the primary user indicator, i.e. $\theta = 1$ implies the presence of the primary user and $\theta = 0$ implies the absence. Given that the primary user transmit with unit power, i.e. $P = 1$, $\sigma^2{}_h$ also represents the received signal power at node $i$ from the primary user. When the signal is received, sensor nodes will do following detection: Given the observation in (1), the detector decides on $H_1 : \theta = 1$ or $H_0 : \theta = 0$ .

### B. *Non-cooperative Spectrum Sensing Strategy*

Now we describe the baseline non-cooperative strategy. In this strategy, every node conducts an energy detection based on the statistics $T(y_i) = |y_i|^2$ [6]. Let $F_0(t)$ and $F_1(t)$ denote the cumulative density function of the random variable $T(y_i)$ under hypothesis $H_0$ and $H_1$ . Since $h_{pi}$ and $w_i$ are both independent complex Gaussian, $T(y_i)$ is exponential distributed with variance of $\theta^2\sigma^2{}_h + \sigma^2{}_w$ . Therefore

$$\begin{aligned} F_0(t) &= P\left(T(y_i) > t \mid H_0\right) = P\left(|w_i|^2 > t\right) \\ &= e^{-t/\left(2\sigma^2{}_w\right)} \end{aligned}, \tag{2}$$

and

$$\begin{aligned} F_1(t) &= P\left(T(y_i) > t \mid H_1\right) = P\left(|h_{pi} + w_i|^2 > t\right) \\ &= e^{-t/\left(2\left(\sigma^2{}_h + \sigma^2{}_w\right)\right)} \end{aligned}. \tag{3}$$

Suppose the predefined maximum false alarm probability is $\alpha$ , then from (2), the corresponding detection threshold $\lambda$ is given by

$$\lambda = -2\sigma^2{}_w \log(\alpha) . \tag{4}$$

And the probability of detection of every sensor node is

$$F_1(\lambda) = \alpha^{\sigma^2{}_w/\left(\sigma^2{}_h + \sigma^2{}_w\right)} = \alpha^{1/(1+SNR)} , \tag{5}$$

where $SNR$ is defined as the ratio of the received useful signal power to noise power. At the fusion center, a majority vote is conducted to decide the presence of the primary user. For $N$ sensors, the probability of detection at the fusion center is

$$p_{fc} = 1 - \sum_{k=0}^{N/2} \binom{N}{k} F_1(\lambda)^k \left(1 - F_1(\lambda)\right)^{N-k} . \tag{6}$$

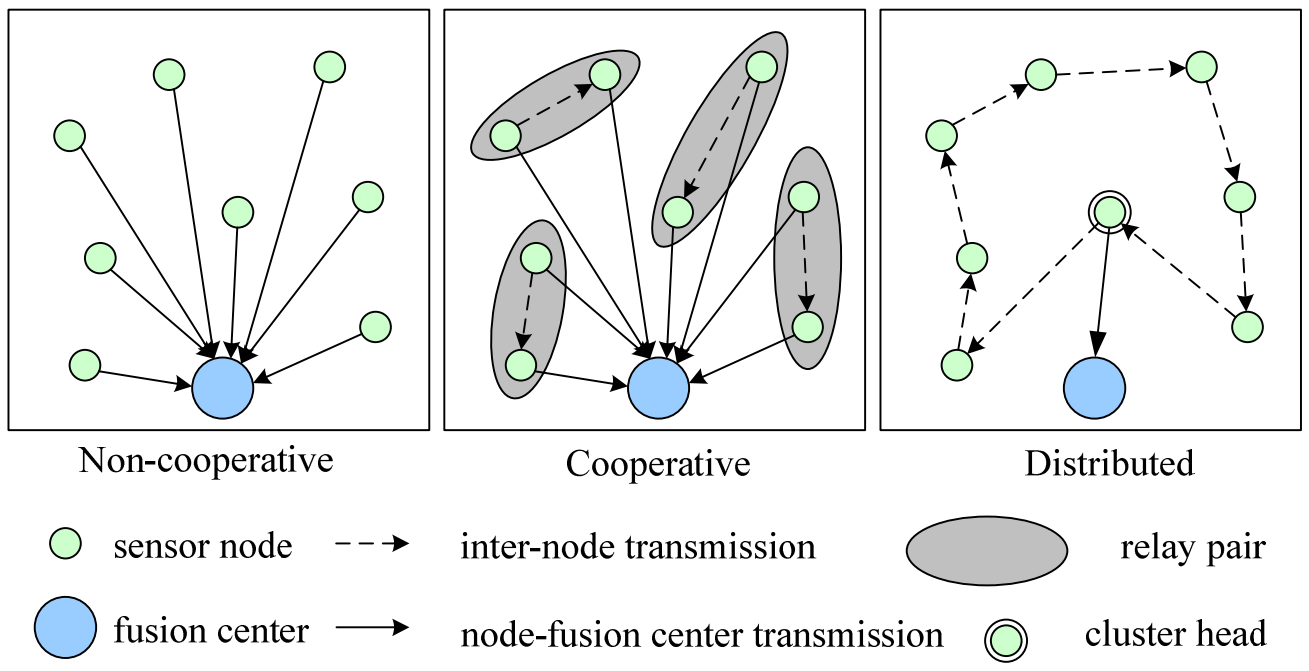

Fig. 1. Spectrum sensing strategies discussed in this paper.

## III. COOPERATIVE SPECTRUM SENSING

The baseline non-cooperative strategy given above has two main disadvantages: 1) Sensor nodes are required to transmit their detection results to the fusion center in every time slot. However, this is unnecessary, since the nodes who fail to detect actually need not to transmit their results. 2) The strategy fails to take advantage of cooperative diversity gain among sensor nodes, which would further increase the detection probability at the fusion center. Taking account of these aspects, we present the following cooperative strategy.

Suppose every two nodes in the sensor network are grouped to form a one-hop relay pair. In the first time slot (T1) of every two consecutive slots, every sensor node makes the same energy detection as in the non-cooperative strategy. If the first node (Node1) in every relay pair fails to detect the primary node, it would amplify and forward (AF) the signal it receives in T1 to the second node (Node2) in its own relay pair in the second time slot (T2). And in T2, Node2 would make the same energy detection as in non-cooperative strategy but with the difference that the signal being observed is the one it received from the primary user plus the one received from Node1. Thus, from [6], the signal received by Node2 in T2 is

$$\begin{aligned} y_2 &= \theta h_{p2} + w_2 + \sqrt{\beta_1} h_{12} y_1 \\ &= \theta h_{p2} + w_2 + \sqrt{\beta_1} h_{12} \left(\theta h_{p1} + w_1\right) \\ &= \left(h_{p2} + \sqrt{\beta_1} h_{12} h_{p1}\right)\theta + w_2 + \sqrt{\beta_1} h_{12} w_1 \end{aligned} , \tag{7}$$

where $\beta_1 \triangleq \tilde{P}/\left(\theta^2\sigma^2{}_h + 1\right)$ is the scaling factor of Node1 in AF, $h_{12}$ is the instantaneous channel gain between Node1 and Node2 within a relay pair, and $y_1$ is the signal received by Node1 in T1. The probability of detection by Node2 is given by [6]

$$p_c = \varphi\left(\lambda;\sigma^2{}_h + 1, \left(\tilde{P}E\left\{\left|h_{12}\right|^2\right\}\left(\sigma^2{}_h + 1\right)\Big/\left(\theta^2\sigma^2{}_h + 1\right)\right)\right), \quad (8)$$

where $\varphi(t;a,b) = \int_0^{\infty} e^{-h-t/(a+bh)}dh$ , $\tilde{P}$ is the maximum relay power constrain, $E\left\{\left|h_{12}\right|^2\right\}$ is the channel gain between the two sensor nodes in a relay pair and $\lambda$ is given by solving equation $\varphi\left(\lambda;1,\tilde{P}E\left\{\left|h_{12}\right|^2\right\}\right) = \alpha$ .

At the fusion center, majority votes are conducted in both T1 and T2. So the probabilities of detection in T1 and T2 are

$$p_{fc_T1} = 1 - \sum_{k=0}^{N/2} \binom{N}{k} F_1(\lambda)^k \left(1 - F_1(\lambda)\right)^{N-k} , \quad (9)$$

and

$$p_{fc_T2} = \left(1 - p_{fc_T1}\right) \cdot \left(1 - \sum_{k=0}^{N'/2} \binom{N'}{k} p_c{}^k \left(1 - p_c\right)^{N'-k}\right), \quad (10)$$

where $N' = \left(1 - F_c(\lambda)\right)N$ is the number of nodes who fail to detect in T1. In all, the probability of detection at the fusion center in both two consecutive time slots is

$$p_{fc} = p_{fc_T1} + p_{fc_T2} . \quad (11)$$

## IV. Primary User Detection Using Distributed Strategy

The strategy proposed above involves cooperation in spectrum sensing among nodes, and accordingly increases the probability of detection. However, the drawback of it also exists. Since both Node1 and Node2 in every relay pair have to transmit their detection results over a *node* $\rightarrow$ *fusion center* distance of $O(1)$, the energy efficiency is not optimal. To this end, we consider a distributed spectrum sensing strategy in sensor network, which reduces the average distance to only $O\left(\sqrt{\log^2 n/n}\right)$ [9]. Notice that estimating the presence of the primary user is like the problem of estimating a series of variables in a sensor network, which can be solved by distributed learning and estimation theory [9][10]. To be specific, build a path through the network which passes through all nodes and visits each node just once. The sequence of nodes can be constructed so that the path hops from neighbor to neighbor. The global detection result can be computed by a single detection process from start node to end node, with each node contributing its own local detection result to the total along the way.

Suppose every sensor node receives its own local signal from the primary user as in (1). Let

$$f\left(\left\{y_i, T_i\right\}_{i\in 1\ldots N}, \theta\right) \triangleq \frac{1}{N}\sum_{i=1}^{N}\left(\left|y_i\right|^2 - T_i\theta\right)^2 , \quad (12)$$

where $T_i$ is the detection threshold. A maximum likelihood estimate for the presence of the primary user is found by solving

$$\hat{\theta} = \arg\min_{\theta} f\left(\left\{y_i, T_i\right\}_{i\in 1\ldots N}, \theta\right) , \quad (13)$$

where $\hat{\theta}$ is the estimation of $\theta$ at the fusion center after iterative computation among nodes. This non-linear least square problem fits well into the general incremental subgradient framework [11]. Taking

$$f_i(\theta) = \left(\left|y_i\right|^2 - T_i\theta\right)^2 , \quad (14)$$

the gradient of $f_i(\theta)$ is

$$\nabla f_i(\theta) = 2\left(\left|y_i\right|^2 - T_i\theta\right)T_i . \quad (15)$$

The theory of incremental subgradient methods shows that given $\nabla f_i(\theta)$ is bounded, $\hat{\theta}$ converges to $\theta$. We assume every sensor node uses the same detection threshold $T_i = T, \forall i$ , thus the gradient is bounded by observing that

$$\left\|\nabla f_i(\theta)\right\| \le 2T\left\|\left|y_i\right|^2 - T\theta\right| < c , \quad (16)$$

where $c$ comes from the assumption on limitation of sensor detection range:

$$\left\|\left|y_i\right|^2 - T\theta\right| < \frac{c}{2T}, \forall i . \quad (17)$$

Now we calculate the probability of detection of this strategy. With $T_i = T, \forall i$ , let $\partial f\left(\left\{y_i, T_i\right\}_{i\in 1\ldots N}, \theta\right)/\partial\theta = 0$ , we get

$$\hat{\theta} = \sum_{i=1}^{N}\left|y_i\right|^2 \Big/ NT . \quad (18)$$

Since $w_i$ is a complex Gaussian, $\left|w_i\right|^2$ and $\sum_{i=1}^{N}\left|w_i\right|^2$ are chi-square distributed variables with 2 and $2N$ degrees of freedom, respectively. Therefore the false alarm probability at the fusion center is given by

$$P\left(\hat{\theta} > t \mid H_0\right) = P\left(\sum_{i=1}^{N}\left|w_i\right|^2 > NTt\right) = e^{-\frac{NTt}{2\sigma^2{}_w}}\sum_{k=0}^{N-1}\frac{1}{k!}\left(\frac{NTt}{2\sigma^2{}_w}\right)^k . \quad (19)$$

With a predefined false alarm probability $\alpha$ , the product $Tt$ in (19) is uniquely determined by solving $P\left(\hat{\theta} > t \mid H_0\right) = \alpha$ since (19) is strictly decreasing in $t$ . Denote $Tt$ as $P'(\alpha)$ , the probability of detection at the fusion center is

$$\begin{aligned} p_d &= P\left(\hat{\theta} > t \mid H_1\right) = e^{-\frac{NTt}{2\left(\sigma^2{}_h + \sigma^2{}_w\right)}}\sum_{k=0}^{N-1}\frac{1}{k!}\left(\frac{NTt}{2\left(\sigma^2{}_h + \sigma^2{}_w\right)}\right)^k \\ &= e^{-\frac{N\cdot P'(\alpha)}{2\left(\sigma^2{}_h + \sigma^2{}_w\right)}}\sum_{k=0}^{N-1}\frac{1}{k!}\left(\frac{N\cdot P'(\alpha)}{2\left(\sigma^2{}_h + \sigma^2{}_w\right)}\right)^k \end{aligned} . \quad (20)$$

## V. Performance Analysis and Discussion

In this section, to compare the non-cooperative, cooperative, and distributed spectrum sensing strategies, three criteria are considered:

- Agility
- Energy efficiency
- Robustness against SNR change

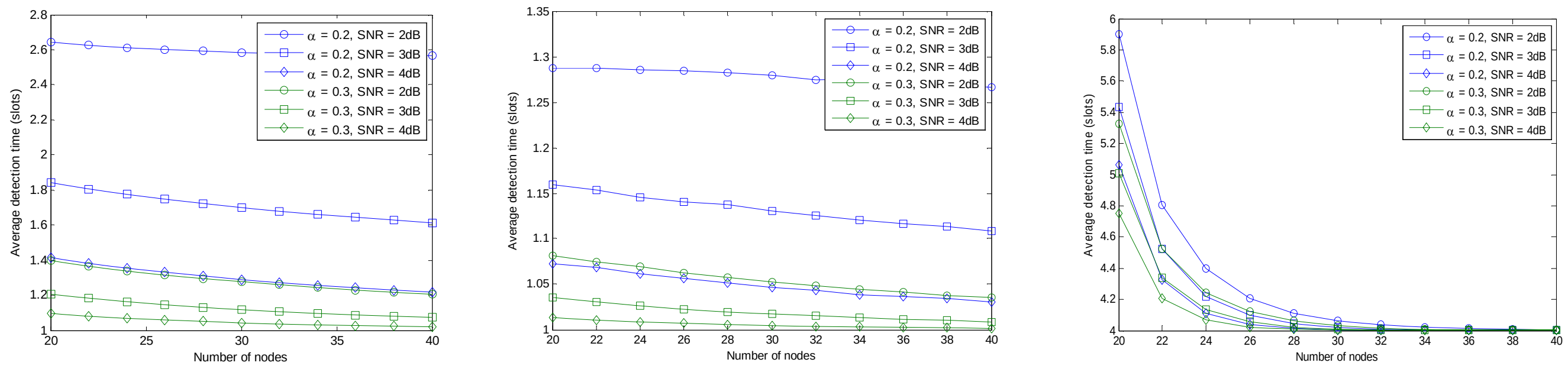

(a) Non-cooperative strategy (b) Cooperative strategy (c) Distributed strategy ( $K = 2$ , $N_s = 2$ )

Fig. 2. Comparison of Average detection time.

### 1) *Agility*

When cognitive users are using the licensed band, they must be able to detect the presence of primary users in a very short time and vacate the band for the primary users as soon as possible. This calls a great agility of detection of the primary users. In this paper, the agility is measured as the number of slots taken by the fusion center to detect the primary user.

Firstly, let us consider the baseline non-cooperative strategy. Let $\tau$ denote the number of slots taken by the fusion center to detect the presence of the primary user, so $\tau$ can be modeled by geometric random distribution as

$$\Pr\{\tau = k\} = \left(1 - p_{fc}\right)^{k-1} p_{fc}, \tag{21}$$

where $p_{fc}$ is the probability of detection by the fusion center from (6). Let $T_{nc}$ be the detection time by the non-cooperative strategy, then

$$T_{nc} = E\{\tau\} = \sum_{k=1}^{+\infty} k\left(1 - p_{fc}\right)^{k-1} p_{fc}. \tag{22}$$

Secondly, in the cooperative strategy, the average detection can be calculated by (22) similarly. But since every detection at the fusion center under the cooperative strategy takes one or two time slots, the result should be multiplied by 1 or 2, i.e.,

$$\begin{aligned} T_c = E\{\tau\} = &\sum_{k=1}^{+\infty} k\left(1 - p_{fc_T1}\right)^{k-1} p_{fc_T1} \\ &+ 2 \cdot \sum_{k=1}^{+\infty} k\left(1 - p_{fc_T2}\right)^{k-1} p_{fc_T2} \end{aligned}, \tag{23}$$

where $p_{fc_T1}$ and $p_{fc_T2}$ are given by (9) and (10). Notice that in this strategy, agility is compensated to involve cooperation among nodes. That is, if the detection at the fusion center in T1 fails, the information of this failed detection would be acquired by the fusion center and be utilized to enhance the detection in T2 by allowing to amplify-and-forward the signal within every relay pair.

Finally, let us consider the distributed strategy. In [12] the authors propose the in-cluster distributed estimation for sensor networks, which greatly reduces latency by a factor of cluster number $N_c$ . Specifically, suppose the whole network is divided into $N_c$ clusters, and each cluster has $N_s \triangleq N / N_c$ sensors. The detections and transmissions in different clusters are conducted simultaneously. In a single iteration, the detection results are transmitted over $N_s - 1$ inter-node hops, and the last cluster head transmits the results to the fusion center. Therefore a single iteration of the distributed strategy takes totally $N_s$ slots. And the total iteration needed to achieve an estimation error smaller than $c^2$ is given by [12]

$$K = \left\lfloor \left\| \hat{\theta}^{(0)} - \theta \right\| / c^2 \right\rfloor, \tag{24}$$

where $c$ is the gradient limitation from (16), and $\hat{\theta}^{(0)}$ is the arbitrary iteration starting point. Therefore, the average detection time in the distributed strategy is given by

$$T_d = N_s \cdot K \cdot \sum_{k=1}^{+\infty} k\left(1 - p_d\right)^{k-1} p_d, \tag{25}$$

where $p_d$ is given by (20).

The comparison of average detection time is in Figure 2, which shows that under the same false alarm probability and SNR, the cooperative strategy achieves the least detection time and the distributed strategy achieves the most. Notice that since the distributed strategy uses at least $K \times N$ time slots, the agility can not be improved further by increasing the node number.

### 2) *Energy Efficiency*

Powered by batteries, sensor nodes in a WSN are very energy-limited, and the main expenditure of energy in a WSN is in the cost of communication. Thus energy efficiency is a critical factor in designing spectrum sensing strategy. In this paper, we mainly focus on two aspects of energy: 1) the total energy consumption needed for a single successful detection at the fusion center and 2) the fairness of energy consumption among sensor nodes. Because the whole longevity of a wireless sensor network is highly affected by the least life time of sensor nodes, fairness of energy consumption is crucial to increase network life time. When considering fairness, we define the fairness degree as the ratio of maximum and minimum energy consumed by nodes in a single time slot, i.e., $\mu \triangleq E_{\max} / E_{\min}$ . Recall that we assume $N$ nodes are uniformly deployed over a square meter, and that the energy cost of every transmission is positive proportional to the transmission distance by a fixed factor $\eta$ . Therefore the average energy consumption of an inter-node transmission is $\eta n^{-1/2}$ , and of a transmission from node to the fusion center is $\eta$ . The total energy used for a single successful detection at the fusion center as a function of the number of nodes is given by $E(n) = e_n \cdot n$, where $n$ is the number of nodes, and $e_n$ is the average amount of energy required to transmit over one hop.

TABLE I
SUMMARY OF PERFORMANCE COMPARISON

| | Non-cooperative | Cooperative | Distributed |
|---|---|---|---|
| **Agility** | Average | Good | Average |
| **Total energy consumption** | Average | Poor (a function of SNR) | Good |
| **Fairness of energy consumption** | Good | Average | Average |
| **Robustness against SNR changes** | Poor | Average | Good |

Firstly, in the non-cooperative strategy, every node needs to transmit its detection result to the fusion center in every slot, therefore $e_n = \eta$ and $E(N) = \eta \cdot N$ . Since $E_{\max} = E_{\min}$ , we have $\mu = 1$.

Secondly, in the cooperative strategy, the energy cost by the first node in every relay pair is $\eta N^{-1/2} + \eta$ and $\eta$ by the second node. So the average total energy consumption is given by

$$\begin{aligned} E(N) &= \eta F_1(\lambda) N + \eta N^{-1/2}(1 - F_1(\lambda)) N + \eta(1 - F_1(\lambda)) N \\ &= (1 - F_1(\lambda)) \eta N^{1/2} + \eta N \end{aligned} \quad .(26)$$

Since $E_{\max} = \eta N^{-1/2} + \eta$, and $E_{\min} = \eta$, so $\mu = 1 + N^{-1/2}$.

Lastly, in the distributed strategy, with a path through which all nodes are visited only once, the distance between neighbor nodes is reduced to $O\left(\sqrt{\log^2 N / N}\right)$ [9]. So in every cluster, except the last cluster head, every node costs $\eta\sqrt{\log^2 N / N}$ , and the cluster head costs $\eta\sqrt{\log^2 N / N} + \eta$. Therefore

$$\begin{aligned} E(N) &= K\left(\eta\sqrt{\log^2 N / N}\,(N - N_c) + \left(\eta\sqrt{\log^2 N / N} + \eta\right) N_c\right) \\ &= KN\eta\sqrt{\log^2 N / N} + KN_c\eta \end{aligned} \quad ,(27)$$

where $K$ is given by (24). Since $E_{\max} = \eta\sqrt{\log^2 N / N} + \eta$ , and $E_{\min} = \eta\sqrt{\log^2 N / N}$ , we have

$$\begin{aligned} \mu &= \left(\eta\sqrt{\log^2 N / N} + \eta\right) \Big/ \left(\eta\sqrt{\log^2 N / N}\right) \\ &= 1 + \sqrt{N} / \log N \end{aligned} \quad . \qquad (28)$$

The performances of the total energy consumption and the fairness are shown in Figure 3 and 4, respectively. It shows that under lower SNRs (such as 0dB), the relation of total energy consumption roughly is: CS ≈ DS<NCS. But since the consumption of the cooperative strategy is a function of $F_1(\lambda)$, hence also a function of the SNR, as the SNR get higher, the energy consumption of the cooperative strategy gets higher as well. Generally speaking, the distributed strategy has the least energy consumption.

Furthermore, Figure 4 shows that the relation of the fairness of energy consumption of the three strategies is: CS ≈ DS< NCS.

*3) Robustness against SNR changes*

In this section, the relation between the detection time and the SNR is studied. In Figure 5, the probabilities of detection at the fusion center of the three strategies are given. It shows that the distributed strategy has a very robust performance against the SNR changes. At lower SNRs, both the non-cooperative and the cooperative strategies have very low probabilities of detection. As the SNR gets higher, the probability of detection of the non-cooperative strategy converges to the one of cooperative strategy. This is because higher SNR leads to higher probability of detection in the first time slot, and thus reduces the contribution of cooperation gain in the second time slot. Moreover, the more the sensor nodes, the faster the convergence goes. This is because the increase of node number acts as a kind of multi-user diversity gain directly. Generally speaking, the relation of the robustness of the three strategies is: DS > CS > NCS.

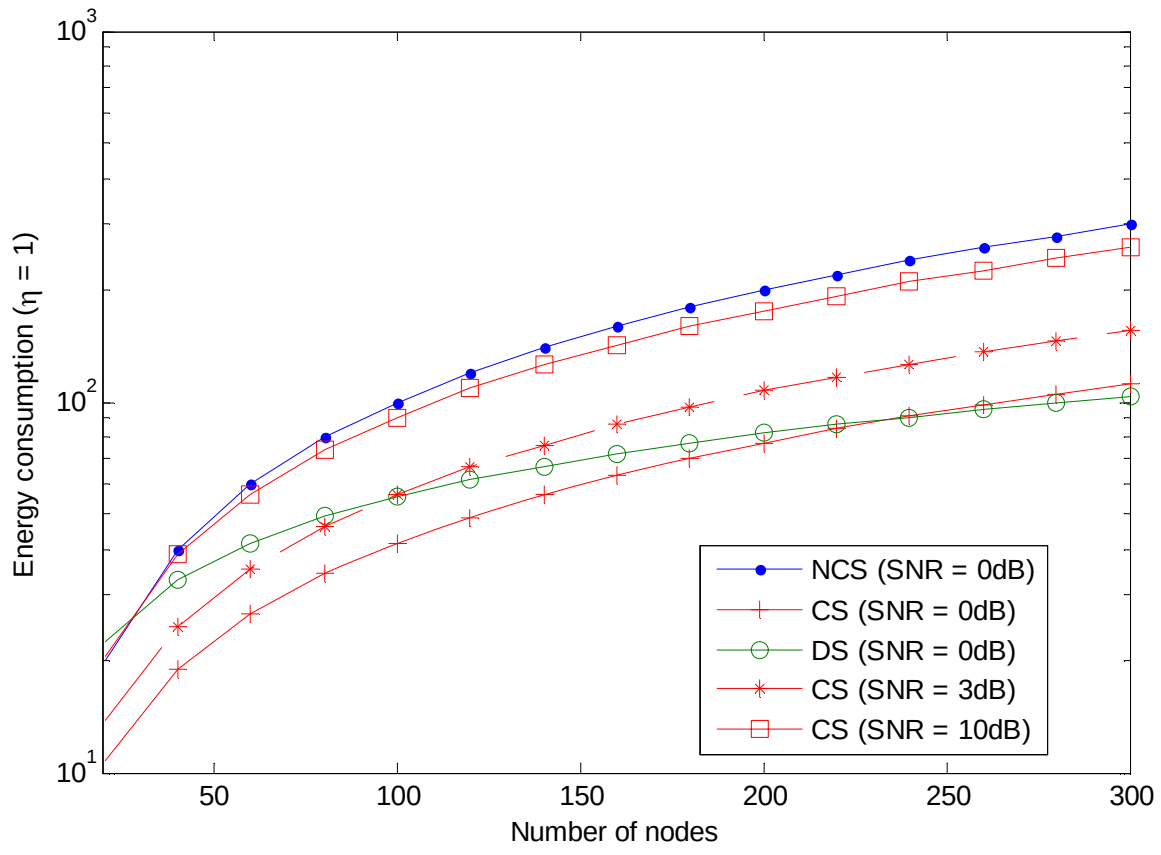

Fig. 3. Comparison of the total energy consumption of the non-cooperative (NCS), the cooperative (CS), and the distributed strategy (DS).

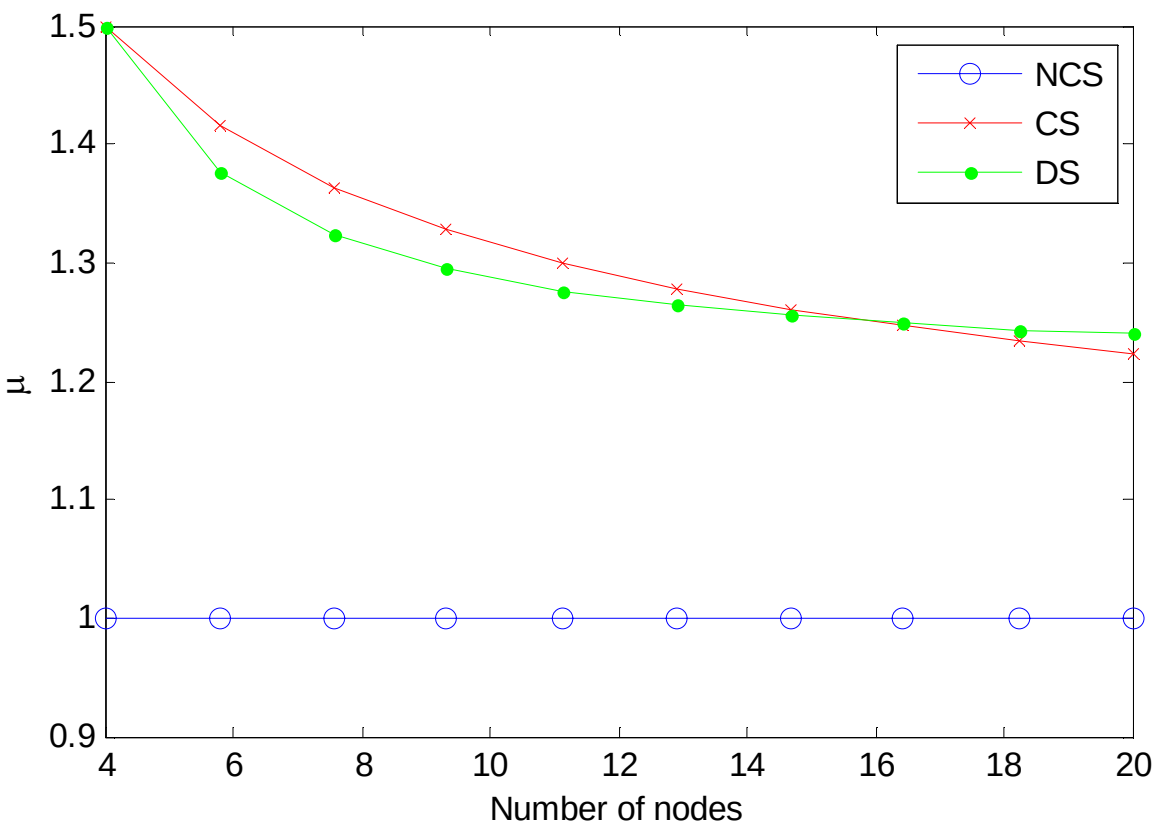

Fig. 4. Comparison of the fairness of energy consumption of the non-cooperative (NCS), the cooperative (CS), and the distributed strategy (DS).

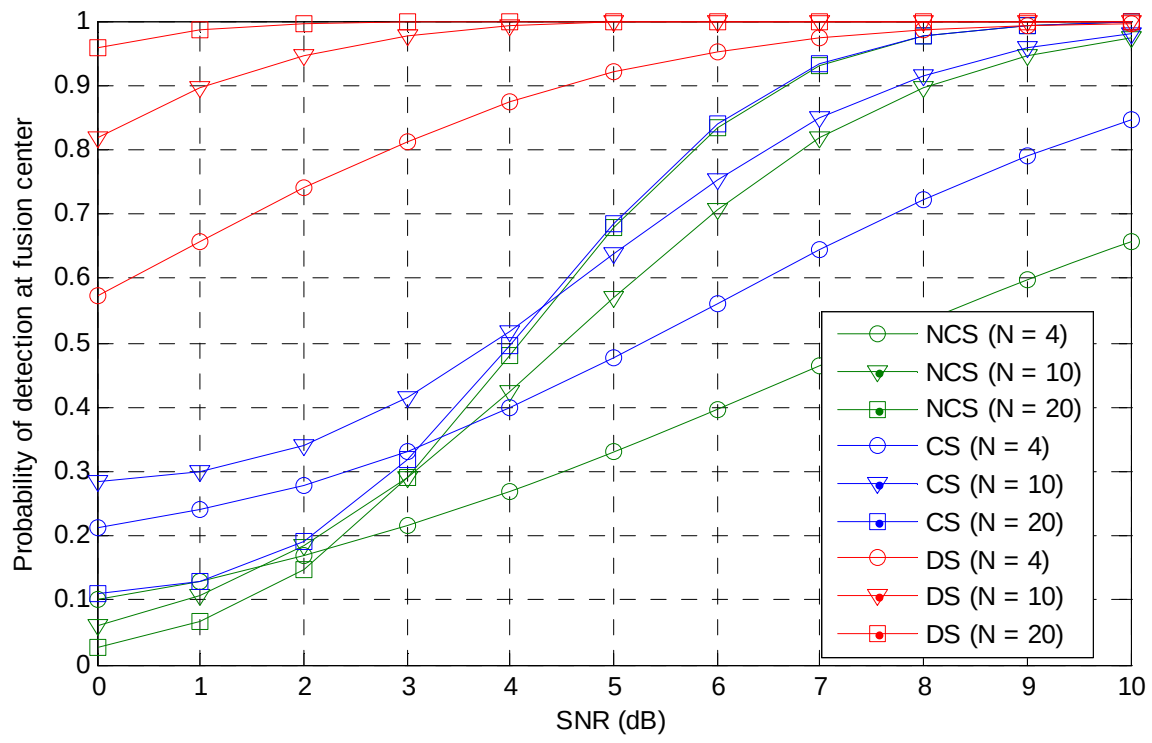

Fig. 5. The probability of detection at fusion center of the non-cooperative (NCS), the cooperative (CS), and the distributed strategy (DS).

The performance comparison is summarized in Table 1.

## VI. Conclusions

In this paper, we focus on the performances of cooperative and distributed spectrum sensing in wireless sensor networks. After introducing a baseline non-cooperative strategy, we have described two strategies: 1) the cooperative strategy, which takes advantage of cooperation diversity gain to increase probability of detection and 2) the distributed strategy, which by passing the results in an inter-node manner increases energy efficiency and fairness among nodes. Analysis shows that the distributed strategy leads to a higher probability of detection at the fusion center than the other two strategies. Furthermore, we have compared the performances of the three strategies based on the criteria of agility, energy efficiency, and the robustness against SNR changes. To sum up, performance comparison shows that: 1) the non-cooperative strategy has the best fairness of energy consumption, 2) the cooperative strategy leads to the best agility, and 3) the distributed strategy leads to the lowest energy consumption and the best robustness against SNR changes.